\pdfoutput=1
\documentclass[11pt]{article}
\usepackage{amsmath}
\usepackage{graphicx}
\usepackage[T1]{fontenc}
\usepackage{times}
\usepackage[top=15mm,bottom=20mm,left=15mm,right=15mm]{geometry}
\usepackage{pdfpages}

\def\clap#1{\hbox to 0pt{\hss#1\hss}}
\def\mathllap{\mathpalette\mathllapinternal}
\def\mathrlap{\mathpalette\mathrlapinternal}
\def\mathclap{\mathpalette\mathclapinternal}
\def\mathllapinternal#1#2{\llap{$\mathsurround=0pt#1{#2}$}}
\def\mathrlapinternal#1#2{\rlap{$\mathsurround=0pt#1{#2}$}}
\def\mathclapinternal#1#2{\clap{$\mathsurround=0pt#1{#2}$}}

\newcommand{\ve}[1]{\mathbf{#1}} 
\newcommand{\ves}[1]{\boldsymbol{#1}} 
\newcommand{\uv}{\widehat{\mathbf{e}}} 
\newcommand{\ma}[1]{#1} 

\newcommand{\prob}[2]{P\!\left({#1}\vphantom{#2}\right\rvert\!\left.\vphantom{#1}{#2}\right)}

\newcommand{\qp}[2]{q_{#1}^+\!\left({#2},{Y}\right)}
\newcommand{\qm}[2]{q_{#1}^-\!\left({#2},{Y}\right)}

\newcommand{\firstn}[1]{\kern0.05em\Gamma{}\kern-0.20em{}_1{}\kern-0.15em{}({#1})}
\newcommand{\secn}[1]{\kern0.05em\Gamma{}\kern-0.20em{}_2{}\kern-0.1em{}({#1})}
\newcommand{\equivn}[1]{\kern0.05em\widetilde{\Gamma}{}\kern-0.20em{}_1{}\kern-0.1em{}({#1})}
\newcommand{\otfn}[1]{\Lambda{}_1{}\kern-0.15em{}({#1})}

\newcommand{\smallsum}[1]{\underset{#1}{\textstyle \sum}}

\newcommand{\tbiglp}{$\smash{\bigl(}$}
\newcommand{\tbigrp}{$\smash{\bigr)}$}

\newcommand{\esmfullstate}{\textsection I}
\newcommand{\esmmotif}{\textsection II}
\newcommand{\esmmotifcoarsegraining}{\textsection II G}
\newcommand{\esmotfdeterministic}{\textsection III}
\newcommand{\esmotfalternative}{\textsection IV}

\begin{document}
\title{Epidemics on contact networks: a general stochastic approach}
\author{Pierre-Andr\'e No\"el, Antoine Allard, Laurent H\'ebert-Dufresne, Vincent Marceau and Louis J. Dub\'e \\ D\'epartement de Physique, de G\'enie Physique et d'Optique, Universit\'e Laval, Qu\'ebec (QC), Canada}
\date{December 8, 2011}
\maketitle
\begin{abstract} 
Dynamics on networks is considered from the perspective of Markov stochastic processes. We partially describe the state of the system through network motifs and infer any missing data using the available information. This versatile approach is especially well adapted for modelling spreading processes and/or population dynamics. In particular, the generality of our systematic framework and the fact that its assumptions are explicitly stated suggests that it could be used as a common ground for comparing existing epidemics models too complex for direct comparison, such as agent-based computer simulations. We provide many examples for the special cases of susceptible-infectious-susceptible (SIS) and susceptible-infectious-removed (SIR) dynamics (\textit{e.g.}, epidemics propagation) and we observe multiple situations where accurate results may be obtained at low computational cost. Our perspective reveals a subtle balance between the complex requirements of a realistic model and its basic assumptions.
\vspace{1ex}
\end{abstract}
%
\section{Introduction}
Mathematical modelling has proven a valuable tool when addressing public health issues. The increase in availability of powerful computer resources has facilitated the use of agent-based models and other complex modelling approaches, all accounting for numerous parameters and assumptions \cite{GLEaMviz2011,Riley2007,Lee2010}. Our confidence in these models may increase when they are shown to agree with empirical observations and/or with previously accepted models. However, when discrepancies appear, the complexity of these computer programs may obfuscate the effect of underlying assumptions, making it difficult to isolate the source of disagreement. While analytical approaches offer more insights on the underlying assumptions, their use is often restricted to simpler interaction structures and/or dynamics.

The purpose of this paper is to systematically model the global behaviour of stochastic systems composed of numerous elements interacting in a complex way. ``Complex'' here implies that interactions among the elements follow some nontrivial patterns that are neither perfectly regular nor completely random, as often seen in real-world systems. ``Stochastic'' implies that the system may not be completely predictable to us and that a probabilistic solution is sought.

To this end, we present (Sec.~\ref{section:generalmodelingscheme}) a general modelling scheme where network theory \cite{barrat_etal_08_book,boccaletti06_pr} accounts for the interactions between the elements of the system and where a birth-death Markov process \cite{gardiner04} models the stochastic dynamics. Since a tremendous amount of information may be required to store the state of the whole system, we seek the part of this information that is important for the problem at hand and then approximate the dynamics by tracking only this limited subset. Part of the discarded data may still affect, albeit weakly, the behaviour of the system. We fill this knowledge gap by inferring the missing information such that it is consistent both with the information we follow and any other prior information that is available to us.

An important part of this paper (Sec.~\ref{section:application}) provides explicit examples to these general ideas. For simplicity, each case either corresponds to a susceptible-infectious-susceptible (SIS) or to a susceptible-infectious-removed (SIR) spreading processes, both standards in the study of infectious diseases propagation. While our first examples study simpler cases, facilitating the understanding of our systematic method, the later models show how the same approach applies to more complex interaction structures.

We then compare and analyse the results of these examples (Sec.~\ref{section:discussion}). This reveals some general considerations for both the accuracy and the complexity of our modelling approach. We find that treating the inferences of missing information explicitly helps systematize the model development and highlights numerous possibilities for future developments. An important simplification occurs for SIR spreading processes and related dynamics, leading to an \emph{exact} model with a small number of dynamical variables.

We conclude (Sec.~\ref{conclusion}) on how our general approach may be applied beyond spreading processes, for example, population dynamics. Returning to the problem of understanding the source of discrepancies in complex models, we explain how \emph{modelling these models} with our method could help identifying important assumptions and isolating the source of disagreement. Mathematical details and further generalizations are also presented in an Electronic Supplementary Material (ESM) \cite{noel11markovESM}.
%
\section{General modelling scheme \label{section:generalmodelingscheme}}
We assume that the real-world system to be modelled is sufficiently well understood to implement a Monte Carlo computer simulation that approximately reproduces its behaviour. We refer to this hypothetical computer simulation as the \emph{full system}: $Z$ denotes the \emph{state} of the full system (\textit{i.e.}, all the data that would be stored by the computer program) while $V$ denotes the \emph{rules} governing the evolution of $Z$ in time (\textit{i.e.}, the program itself).

However, there are many situations where a direct implementation of the full system is impractical due to storage and/or computation considerations. We thus design a \emph{simplified system} that aims at reproducing the behaviour of the full one, while requiring less resources.

The state $X$ of this simplified system (\emph{much} smaller than $Z$) evolves in time according to the rules $W$. Moreover, we note $Y$ any known \emph{prior information} that is relevant in a Bayesian inference of $Z$
\begin{equation}
  \prob{Z}{X,Y} = \frac{ \prob{Z}{Y} \prob{X}{Y,Z} }{ \prob{X}{Y} } \quad .
\end{equation}
This last point is crucial: $Y$ often makes the difference between an accurate model and a useless one. It bridges the gap between the simplified representation of the state of the system (\textit{i.e.}, $X$) and the full one (\textit{i.e.}, $Z$).

Since we are interested in systems composed of many elements interacting through complex patterns, we express the previous quantities in terms of networks.
%
\subsection{Networks \label{subsection:networks}}
A \emph{network} (graph) is a collection of \emph{nodes} (vertices) and \emph{links} (edges). Nodes model the elements of a system; links join nodes pairwise to represent interactions between the corresponding elements. Two nodes sharing a link are said to be \emph{neighbours} and the \emph{degree} of a node is its number of neighbours. The part of a link that is attached to a node is called a \emph{stub}: there are two stubs per link and each node is attached to a number of stubs equal to its degree. A link with both ends leading to the same node is called a \emph{self-loop} and \emph{repeated links} occur when more than one link join the same two nodes.

There may be systems such that specifying its state $Z$ exactly amounts to specifying the network structure. However, most systems are not purely structural: they are composed of elements that, by themselves, require additional information to be properly characterized. Hence, we assign to each node a \emph{node state} that specifies the intrinsic properties of the corresponding element in the system. Both the structure and these intrinsic node states are specified by $Z$; see ESM \cite{noel11markovESM} \esmfullstate\ for further examples of information that may be contained in $Z$, including the important case of directed networks.
%
\subsection{Motifs \label{subsection:motifs}}
Specifying the complete structure of a complex network requires a tremendous amount of information. Since we want the state $X$ of a simplified system to be of manageable size, approximations have to be made. A convenient way to do so, and one that has proven to give good results in the past \cite{house2009,karrer10,Gleeson11PRL,Marceau2010pre,HebertDufresne2010pre,Marceau2011pre,noel11otf}, is to specify the network structure through its building blocks.

A network \emph{motif} is a pattern that may appear a number of times in the network. For example, two linked nodes form a \emph{pair motif} while three nodes all neighbours of one another form a \emph{triangle motif}. Motifs may encode intrinsic node states or other relevant information; further details and examples are provided throughout Section~\ref{section:application} as well as in ESM \cite{noel11markovESM} \esmmotif.

We define the \emph{state vector} $\ve{x}$ of a system as a vector of integers specifying how many times different motifs appear in the network. These motifs may be attached together to form a network structure: the state vector $X = \ve{x}$ enumerates the available building blocks while the prior information $Y$ specifies how such blocks may be attached. There will usually be numerous valid ways to attach the blocks, some more probable than others. Given the available information, the resulting distribution is our best estimate for $\prob{Z}{\ve{x}, Y}$.

By judiciously choosing the motifs enumerated in $\ve{x}$ and by specifying informative prior information $Y$, one may hope for this probability distribution to be densely localized around the ``real'' value of $Z$ in the full system. This mapping can then be used to convert the rules $V$ of the full system to the rules $W$ of the new simplified one. We approach this problem from the perspective of birth-death Markov processes.
%
\subsection{Birth-death stochastic processes \label{subsection:birthdeath}}
In a \emph{birth-death} process, the elements composing a system may be destroyed (death) while new ones may be created (birth). It is therefore natural to state the rules $W$ of our simplified system in those terms: any change in the state vector $\ve{x}$ may be perceived as an event where motifs are created and/or destroyed.

Quantitatively, a \emph{forward transition event of type $j$} takes the system from state $\ve{x}$ to state $\ve{x} + \ve{r}^j$ and has probability $\qp{j}{\ve{x}} dt$ to occur during the time interval $[t,t+dt)$. Similarly, a \emph{backward transition event of type $j$} takes the system from state $\ve{x}$ to state $\ve{x} - \ve{r}^j$ and has probability $\qm{j}{\ve{x}} dt$ to occur during the same time interval.

Specifying for each $j$ the elements $r_i^j$ of the \emph{shift vector} $\ve{r}^j$ together with the \emph{rate functions} $\qp{j}{\ve{x}}$ and $\qm{j}{\ve{x}}$ thus completely define the rules $W$ governing the simplified system. This Markov process is summarized in the master equation
\begin{align}
  \frac{d \prob{\ve{x}}{Y,t}}{dt} & = \sum_j \biggl[ \qp{j}{\ve{x} - \ve{r}^j} \prob{\ve{x} - \ve{r}^j}{Y,t} - \qp{j}{\ve{x}} \prob{\ve{x}}{Y,t} \label{eq:master} \\
  & \qquad + \qm{j}{\ve{x} + \ve{r}^j} \prob{\ve{x} + \ve{r}^j}{Y,t} - \qm{j}{\ve{x}} \prob{\ve{x}}{Y,t} \biggr] \nonumber
\end{align}
specifying the evolution of the probability $\prob{\ve{x}}{Y,t}$ to observe state $\ve{x}$ at time $t$ (notation compatible with \cite{gardiner04} \textsection 7.5).

We now consider two approximations that are often justified for large systems: the elements of $\ve{x}$ may be treated as varying continuously and the probability distribution is strongly concentrated around its mean value. In such cases, the evolution of the mean value $\ves{\mu}(t) = \sum_{\ve{x}} \ve{x} \prob{\ve{x}}{Y,t}$ for the vector $\ve{x}$ at time $t$ is approximately given by
\begin{subequations}
\begin{gather}
  \frac{d \ves{\mu}(t)}{dt} = \ve{a}\bigl( \ves{\mu}(t) \bigr) \label{eq:ApproxMean:dmudt} \\
  a_i(\ve{x}) = \sum_j r_i^j \left[ \qp{j}{\ve{x}} - \qm{j}{\ve{x}} \right] \label{eq:ApproxMean:a}
\end{gather} \label{eq:ApproxMean}
\end{subequations}
where we defined the \emph{drift vector} $\ve{a}(\ve{x})$ of elements $a_i(\ve{x})$ (see \cite{gardiner04} \textsection 7.5.3 and \textsection 4.4.9).

In order to further refine our knowledge of $\prob{\ve{x}}{Y,t}$ in the vicinity of this deterministic solution, we define the \emph{evolution matrix} $\ma{A}(t,t')$, the \emph{diffusion matrix} $\ma{B}(\ve{x})$ \tbiglp of elements $B_{ii'}(\ve{x})$\tbigrp and the \emph{covariance matrix} $\ma{C}(t)$
\begin{subequations} \label{eq:GaussianApprox}
\begin{align}
  \ma{A}(t,t') & = \exp\left[\int_{t'}^t \ma{J}_{\ve{a}}\bigl( \ves{\mu}(t'') \bigr) dt''\right] \\
  B_{ii'}(\ve{x}) & = \sum_j r_i^j r_{i'}^j \Bigl[ \qp{j}{\ve{x}} + \qm{j}{\ve{x}} \Bigr] \\
  \ma{C}(t) & = \int_0^t \ma{A}(t,t') \cdot \ma{B}\bigl( \ves{\mu}(t') \bigr) \cdot \ma{A}(t,t')^T dt'
\end{align}
\end{subequations}
where $\ma{J}_{\ve{a}}(\ve{x})$ is the Jacobian matrix of $\ve{a}$ evaluated at $\ve{x}$. Noting $d$ the size of the vector $\ve{x}$, the probability distribution may be approximated by a $d$-dimensional Gaussian
\begin{equation}
  \prob{\ve{x}}{Y,t} = \frac{ \exp \left\{ \! -\frac{1}{2} \bigl[\ve{x}(t) - \ves{\mu}(t) \bigr]^T \! \! \! \cdot \! \ma{C}(t)^{-1} \! \cdot \! \bigl[\ve{x}(t) - \ves{\mu}(t) \bigr] \! \right\} }{\sqrt{(2\pi)^d \bigl\lvert \ma{C}(t) \bigr\rvert}} \label{eq:dGaussian}
\end{equation}
where $\bigl\lvert \ma{C}(t) \bigr\rvert$ is the determinant of $\ma{C}(t)$. Note that \eqref{eq:master}--\eqref{eq:dGaussian} are all textbook relationships.

Although many other tools are available for the analysis of stochastic systems, the simplicity, the generality and the straightforwardness of the Gaussian approximation make it an instrument of choice that will be used extensively in this article.
%
\section{Application to spreading dynamics \label{section:application}}
Without prejudice to the generality of Section~\ref{section:generalmodelingscheme}, we now focus our study to spreading processes \cite{keeling05_jrsi,bansal07_jrsi,Danon2010review}. An epidemiological terminology is used: whatever propagates among neighbouring nodes, be it desirable or not, is called an \emph{infection}. We find that the basic SIS and SIR epidemiological models, both to be defined shortly, require little prior knowledge from the part of the reader while being sufficiently complex for the needs of the present study.

At a given time, the intrinsic state of each node of an \emph{SIS model} may either be \emph{Susceptible} (not carrying the infection) or \emph{Infectious} (carrying the infection). The full system state $Z$ hence specifies each node's intrinsic state together with the complete structure of the network. The rules $V$ are simple: during any time interval $[t,t+dt)$, each infectious node may recover (\textit{i.e.}, it becomes susceptible) with probability $\alpha\, dt$ and, for each of its susceptible neighbours, has probability $\beta\, dt$ to transmit the infection (\textit{i.e.}, the neighbour becomes infectious).

In addition to the susceptible and infectious intrinsic states, the nodes of an \emph{SIR model} may also be \emph{Removed} (once had the infection and can neither acquire nor transmit it ever again). The rules $V$ are the same than for the SIS model with respect to infection (\textit{i.e.}, infectious nodes transmit to their susceptible neighbour with probability $\beta\, dt$), but recovery is replaced by removal (\textit{i.e.}, infectious nodes become removed with probability $\alpha\, dt$).

The remainder of this section studies how different choices of state vector $\ve{x}$ and prior information $Y$ translate in the rules $W$ of the simplified system. In each case, $W$ is defined through a set of equations whose tags all share the same numeral, \textit{e.g.}, \eqref{eq:SISpairRegular:x}--\eqref{eq:SISpairRegular:inferenceS}. Although figures present results concomitantly with the specification of the corresponding models, all discussions are delayed to Section~\ref{section:discussion}.
%
\subsection{Pair-based SIS model \label{subsection:pairSIS}}
Section~\ref{subsection:motifs} defined a pair motif as two linked nodes. Since the nodes of a SIS model are either susceptible or infectious, there are three possibilities for pair motifs: two linked susceptible nodes (noted $S\text{--}S$), two linked infectious nodes (noted $I\text{--}I$) and a susceptible node linked to an infectious one (noted $S\text{--}I$). Two nodes involved in a pair motif may have other neighbours.

Pair motifs are often used in conjunction with \emph{node motifs}: the trivial structure that is one node. In the SIS model, there are two possibilities for a node motif: susceptible nodes (noted $S$) and infectious nodes (noted $I$). A state vector $\ve{x}$ based on both node and pair motifs would thus be composed of five elements enumerating the amount of times each motif appears in the network: $x_S$, $x_I$, $x_{S\text{--}S}$, $x_{S\text{--}I}$ and $x_{I\text{--}I}$. However, additional assumptions about the structure of the network may cause some of these quantities to be redundant.
%
\subsubsection{Degree-regular network \label{subsubsection:degreeregular}}
We first consider the simple case where the network is known to be a \emph{$n$-regular network} of size $N$: there are $N$ nodes in the network which all have $n$ neighbours (degree $n$). Such a network must respect the structural constraints $x_S = N - x_I$, $x_{S\text{--}S} = \frac{1}{2} \left( n x_S - x_{S\text{--}I} \right)$ and $x_{I\text{--}I} = \frac{1}{2} \left( n x_I - x_{S\text{--}I} \right)$. Hence, with the prior information $Y$ specifying $N$ and $n$, the state vector
\begin{subequations} \label{eq:SISpairRegular}
\begin{equation}
  \ve{x} = \bigl( x_I, x_{S\text{--}I} \bigr)  \label{eq:SISpairRegular:x}
\end{equation}
suffices to obtain all the five node and pair motifs.

In those terms, the rules $V$ specify that an infection has probability $\beta\, x_{S\text{--}I}\, dt$ to occur during the time interval $[t,t+dt)$ while a recovery has probability $\alpha\, x_I\, dt$ to occur. Clearly, an infection translates to the destruction of a $S$ motif and the creation of a new $I$ one, and a recovery corresponds to the inverse process. However, pair motifs are also affected by such transitions since the affected node had neighbours. Hence, the effect on $\ve{x}$ of the infection or recovery of a node depends on some information that is not directly tracked by $\ve{x}$ --- \textit{i.e.}, what is the state of the infected or recovered node's neighbours --- and we thus have to \emph{infer} this information from the available data.

In order to facilitate this inference, we define the \emph{first neighbourhood motif} $S \firstn{k_S,k_I}$ as a susceptible node that has $k_S$ susceptible neighbours and $k_I$ infectious neighbours. Similarly, the motif $I \firstn{k_S,k_I}$ corresponds to an infectious node with $k_S$ susceptible neighbours and $k_I$ infectious ones. In both cases, we qualify as \emph{central} the node whose neighbours are explicitly stated. The other nodes of the first neighbourhood motif, \textit{i.e.}, the neighbours of the central node, may have other neighbours of their own.

We can now define a forward transition event of type $j \in \{ 0, 1, \ldots, n \}$ as the infection of the central node of a $S \firstn{n-j,j}$ motif. In terms of node and pair motifs, this implies the destruction of one of the $S$ motifs, of $n-j$ of the $x_{S\text{--}S}$ motifs and of $j$ of the $x_{S\text{--}I}$ motifs together with the creation of one new $I$ motif, of $n-j$ new $x_{S\text{--}I}$ motifs and of $j$ new $x_{I\text{--}I}$ motifs. Since only $x_I$ and $x_{S\text{--}I}$ are tracked, the shift vectors are
\begin{equation}
  \ve{r}^j = \bigl( 1, n - 2j \bigr) \quad . \label{eq:SISpairRegular:r}
\end{equation}
This same vector also defines the backward transition events $j \in \{ 0, 1, \ldots, n \}$ which correspond to the recovery of the central node of a $I \firstn{n-j,j}$ motif.

Looking back at the rules $V$, the corresponding forward and backward transition rate functions are
\begin{align}
  \!\qp{j}{\ve{x}} & = \beta\, x_{S\text{--}I}\, \prob{S \firstn{n-j,j}}{S,j \ge 1,\ve{x},Y} \label{eq:SISpairRegular:qp} \\
  \!\qm{j}{\ve{x}} & = \alpha\, x_I\, \prob{I \firstn{n-j,j}}{I,\ve{x},Y} \label{eq:SISpairRegular:qm}
\end{align}
where two inference terms have been defined.

The inference term of \eqref{eq:SISpairRegular:qm} gives the probability for a motif to be a $I \firstn{n-j,j}$ knowing that it has an infectious node at its center and that the current state vector is $\ve{x}$ with prior information $Y$. For a sufficiently large network, this approximately corresponds to randomly drawing the $n$ neighbours of the central infectious node among the pair motifs $S\text{--}I$ and $I\text{--}I$
\begin{equation}
  \prob{I \firstn{n-j,j}}{I,\ve{x},Y} = \binom{n}{j} \left( \frac{x_{S\text{--}I}}{n\, x_I} \right)^j \! \left( 1 - \frac{x_{S\text{--}I}}{n\, x_I} \right)^{n-j} \label{eq:SISpairRegular:inferenceI} \quad .
\end{equation}
The inference term of \eqref{eq:SISpairRegular:qp} is very similar except that the central susceptible node is known to have at least one infectious neighbours since it acquired the infection through a $S\text{--}I$ motif
\begin{equation}
  \prob{S \firstn{n-j,j}}{S,j \ge 1,\ve{x},Y} = \binom{n-1}{j} \left( \frac{x_{S\text{--}I}}{n\, (N - x_I)} \right)^j \! \left( 1 - \frac{x_{S\text{--}I}}{n\, (N - x_I)} \right)^{n-1-j} \label{eq:SISpairRegular:inferenceS}
\end{equation}
\end{subequations}
which complete the rules $W$ for the pair-based SIS model on a $n$-regular network of size $N$.

\begin{figure}
\begin{center}
\includegraphics{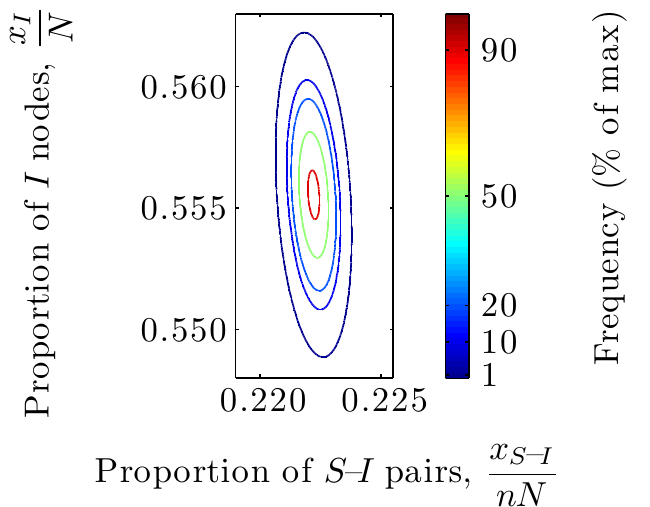}
\end{center}
\caption{(Online version in colour.) Distribution of post-transient ($t \rightarrow \infty$) outcomes as predicted by the SIS model on regular network \eqref{eq:SISpairRegular} using the approximations \eqref{eq:ApproxMean}--\eqref{eq:GaussianApprox}. The axes (proportion of $I$ among node motifs \textit{vs} proportion of $S\text{--}I$ among pair motifs), network structure ($N = 10^5$ nodes, each of degree $n = 5$) and parameters ($\alpha = 0.1$\protect\footnote{M. J. Keeling confirmed us that $\alpha=0.01$ in \protect\cite{dangerfield09} (noted $\gamma$) is a typo.} and $\beta = 0.05$) are the same as for Fig.~2(c) from \cite{dangerfield09}. Frequencies (in percent) are used to facilitate comparison with \cite{dangerfield09}; they are simply obtained from the probability densities of \eqref{eq:dGaussian} multiplied by $100$. \label{fig:SISpairRegular:dist}}
\end{figure}
\begin{figure}
\begin{center}
\includegraphics{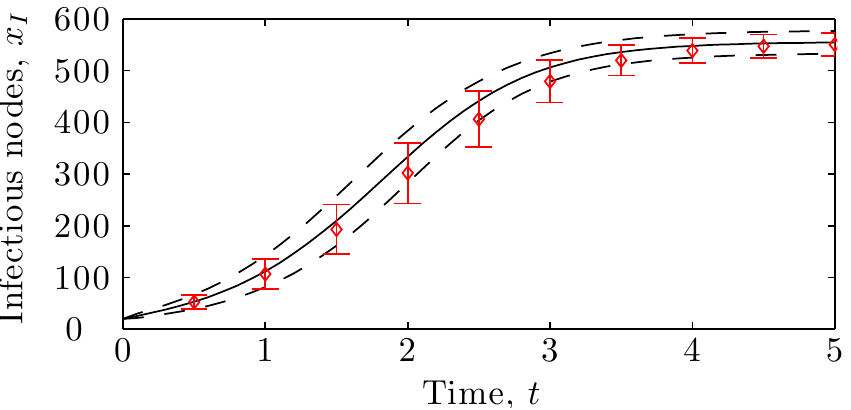}
\end{center}
\caption{(Online version in colour.) Time evolution of the number of infectious nodes $x_I$ for SIS dynamics ($\alpha = 2$ and $\beta = 1$) on a regular network of $N = 10^3$ nodes ($20$ initially infectious) of degree $n = 5$. Curves: results for the simplified system \eqref{eq:SISpairRegular} approximated with \eqref{eq:ApproxMean}--\eqref{eq:GaussianApprox}. The continuous curve shows the mean value while the dashed curves delimit the range of one standard deviation above and below the mean. Symbols: averaged results of $10^5$ numerical simulations of the full system. The parameters $\alpha$ and $\beta$ correspond to those of Fig.~\ref{fig:SISpairRegular:dist} after rescaling the time unit. \label{fig:SISpairRegular:time}}
\end{figure}
\begin{figure}
\begin{center}
\includegraphics{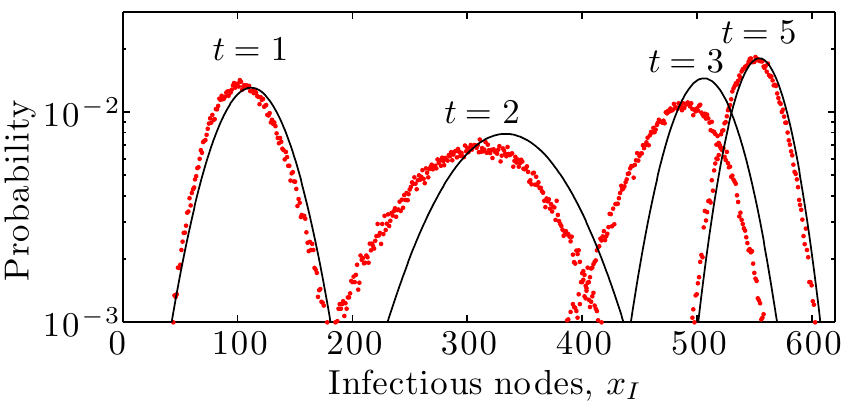}
\end{center}
\caption{(Online version in colour.) Probability distribution at different times for the number of infectious nodes $x_I$. All parameters are the same as in Fig.~\ref{fig:SISpairRegular:time}. Curves: Gaussian approximation for the simplified system. Symbols: binned results of $10^5$ numerical simulations of the full system. \label{fig:SISpairRegular:timeslices}}
\end{figure}

Figure~\ref{fig:SISpairRegular:time} compares the results produced by this simplified model (defined by $W$, $\ve{x}$ and $Y$) to the corresponding full one (defined by $V$ and $Z$). Figure~\ref{fig:SISpairRegular:timeslices} shows the probability distribution for the same data. Note that, although presented differently, this model corresponds to the one presented in \cite{dangerfield09}; Fig.~\ref{fig:SISpairRegular:dist} is provided for comparison with Fig.~2(c) of \cite{dangerfield09}.
%
\subsubsection{Erd\H{o}s-R\'enyi network \label{subsubsection:SISpairER}}
We now consider the case where the network is an Erd\H{o}s-R\'enyi network: there are $N$ nodes in the networks and $M$ links are randomly assigned. This knowledge constrains two of the five node and pair motifs (\textit{i.e.}, $x_S = N - x_I$ and $x_{S\text{--}S} = M - x_{S\text{--}I} - x_{I\text{--}I}$) and a state vector of three elements suffices
\begin{subequations} \label{eq:SISpairER}
\begin{equation}
  \ve{x} = \bigl( x_I, x_{S\text{--}I}, x_{I\text{--}I} \bigr) \quad . \label{eq:SISpairER:x}
\end{equation}

The method used in Section~\ref{subsubsection:degreeregular} has to be adapted since the degree of each node is not constrained to a single value. Indeed, a susceptible node that gets infected may \emph{a priori} be the center of any of the $S \firstn{k_S,k_I}$ motifs. Still, we could design a bijective mapping between the vector of integers $\ve{k} = (k_S,k_I)$ and an event type $j$.

The details of the chosen mapping do not matter: we simply define the forward transition event of type $\ve{k}$ as the infection of the central node of a $S \firstn{k_S,k_I}$ motif, which may conveniently be noted $S \firstn{\ve{k}}$ instead. Similarly, the backward transition event of type $\ve{k}$ is defined as the recovery of the central node to a $I \firstn{\ve{k}}$ motif. The corresponding shift vector and rate functions are
\begin{align}
  \ve{r}^{\ve{k}} & = \bigl( 1, k_S - k_I, k_I \bigr) \label{eq:SISpairER:r} \\
  \qp{\ve{k}}{\ve{x}} & = \beta\, x_{S\text{--}I}\, \prob{S \firstn{\ve{k}}}{S,k_I \! \ge \! 1,\ve{x},Y} \label{eq:SISpairER:qp} \\
  \qm{\ve{k}}{\ve{x}} & = \alpha\, x_I\, \prob{I \firstn{\ve{k}}}{I,\ve{x},Y} \label{eq:SISpairER:qm}
\end{align}
where the inference terms bear the same meaning as their previous counterpart. Again assuming large network size, these inference terms are obtained by evaluating the probability for each pair motif to include the considered node. Hence, the products of binomial distributions
\begin{align}
  \prob{S \firstn{\ve{k}}}{S,l \ge 1,\ve{x},Y} & = \binom{2\, x_{S\text{--}S}}{k_S} x_S^{-k_S} \bigl(1-x_S^{-1}\bigr)^{2\, x_{S\text{--}S} - k_S} \binom{x_{S\text{--}I}-1}{k_I-1} x_S^{-(k_I-1)} \bigl(1-x_S^{-1}\bigr)^{x_{S\text{--}I} - k_I} \label{eq:SISpairER:inferenceS} \\
  \prob{I \firstn{\ve{k}}}{I,\ve{x},Y} & = \binom{x_{S\text{--}I}}{k_S} x_I^{-k_S} \bigl(1-x_I^{-1}\bigr)^{x_{S\text{--}I} - k_S} \binom{2\, x_{I\text{--}I}}{k_I} x_I^{-k_I} \bigl(1-x_I^{-1}\bigr)^{2\, x_{I\text{--}I} - k_I} \label{eq:SISpairER:inferenceI}
\end{align}
\end{subequations}
complete the rules $W$ for the pair-based SIS model on a Erd\H{o}s-R\'enyi network of size $N$ with $M$ links.

\begin{figure}
\begin{center}
\includegraphics{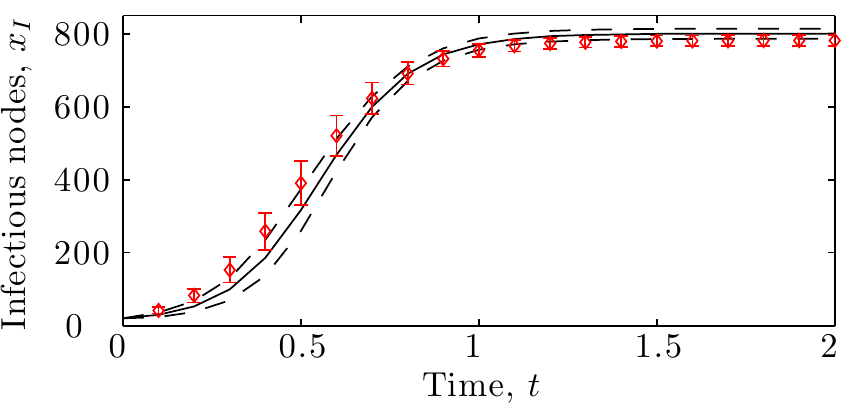}
\end{center}
\caption{(Online version in colour.) Time evolution of the number of infectious nodes $x_I$ for SIS dynamics ($\alpha = 2$ and $\beta = 1$) on an Erd\H{o}s-R\'enyi network of $N = 10^3$ nodes ($20$ initially infectious) and $M = 5 \cdot 10^3$ links. The mean and range of one standard deviation above and bellow the mean are shown. Curves: simplified system. Symbols: full system ($10^5$ simulations). \label{fig:SISpairER:time}}
\end{figure}

Figure~\ref{fig:SISpairER:time} compares the results produced by this simplified model to the corresponding full one. Section~\ref{subsection:discussion:pair} discusses these results and provides further details concerning pair-based models.
%
\subsection{First neighbourhood SIS model \label{subsection:SISfirst}}
We consider a full model ($V$ and $Z$) for SIS dynamics on a \emph{configuration model} (CM) network: given a sequence $\{n_0, n_1, n_2, \cdots\}$, links are randomly assigned between nodes such that, for each degree $\kappa$, there are $n_\kappa$ nodes of degree $\kappa$. In a computer simulation, we create $n_\kappa$ nodes with $\kappa$ stubs for each possible $\kappa$ and then randomly pair stubs to form links. No particular mechanism is used to prevent the formation of repeated links and self-loops: this simplifies the analytical treatment and has little effect when the network size is sufficiently large.

Our simplified model handles the heterogeneity in node degree by enumerating every possible first neighbourhood motifs in its state vector
\begin{subequations} \label{eq:SISfirst}
\begin{equation}
  \ve{x} = \bigl( x_{S\firstn{0,0}}, x_{S\firstn{1,0}}, \cdots, x_{I\firstn{0,0}}, \cdots \bigr) \quad . \label{eq:SISfirst:x}
\end{equation}
Although this vector should be infinite in the general case, it is not the case when, \textit{e.g.}, the prior information $Y$ states that no node has a degree superior to $\mathcal{K}$.

For the same reasons that models tracking node and pair motifs (Section~\ref{subsection:pairSIS}) had their transition events defined in terms of first neighbourhood motifs, the transition events are here defined in terms of \emph{second neighbourhood motifs}: a central node, its neighbours and the neighbours of those neighbours. In the same way that we note $\nu\firstn{\ve{k}}$ the first neighbourhood motif formed by a state $\nu$ central node with neighbourhood specified by $\ve{k}$, we note $\nu\secn{\ve{K}}$ the second neighbourhood motif formed by a state $\nu$ central node with neighbourhood specified by $\ve{K}$. 

The elements of $\ve{K}$ may be indexed with first neighbourhood motifs: the central node has $K_{\nu'\firstn{\ve{k}'}}$ state $\nu'$ neighbours whose other neighbours (\textit{i.e.}, excluding the central node) are specified by $\ve{k}'$. Hence, the second neighbourhood motif
\begin{center}
\vspace{-0.5\bigskipamount}
\includegraphics{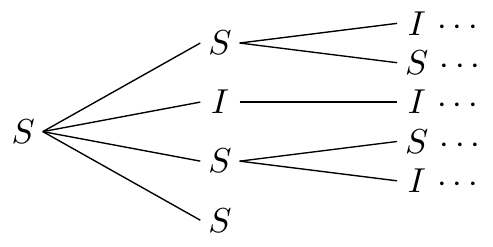}
\vspace{-0.5\bigskipamount}
\end{center}
is noted $S\secn{\ve{K}}$ with all elements of $\ve{K}$ zero except for $K_{S\firstn{0,0}} = 1$, $K_{I\firstn{0,1}} = 1$ and $K_{S\firstn{1,1}} = 2$. Note that the central node of this second neighbourhood motif is also the central node of the first neighbourhood motif $S\firstn{3,1}$. In general, we note $\nu\equivn{\ve{K}}$ the first neighbourhood motif that shares the same central node as the second neighbourhood motif $\nu\secn{\ve{K}}$.

We digress further to introduce the \emph{unit vector} notation $\uv_{\mathcal{M}}$ where $\mathcal{M}$ represents a motif; all the elements of this vector are zero except for the $\mathcal{M}$-th, which is one. The total number of elements in $\uv_{\mathcal{M}}$ should be clear from the context. As a concrete example, the right hand side of \eqref{eq:SISpairRegular:r} could be noted $\uv_I + (n-2j)\uv_{S\text{--}I}$.

Similarly to Section~\ref{subsubsection:SISpairER}, we define the forward transition event of type $\ve{K}$ to be the infection of the central node of a $S\secn{\ve{K}}$ motif and the backward transition event of type $\ve{K}$ as the recovery of the central node of a $I\secn{\ve{K}}$ motif. The corresponding shift vector is
\begin{equation}
  \ve{r}^{\ve{K}} = \uv_{I\equivn{\ve{K}}} - \uv_{S\equivn{\ve{K}}} + \sum_\nu \sum_{\ve{k}} K_{\nu\firstn{\ve{k}}} \bigl( \uv_{\nu\firstn{\ve{k} + \uv_I}} - \uv_{\nu\firstn{\ve{k} + \uv_S}} \bigr) \quad . \label{eq:SISfirst:r}
\end{equation}
The first line shows the direct effect of a change of state in the central node while the second one handles the ``collateral effect'' on its immediate neighbours. Here the unit vector $\uv_\nu$ has the same dimension as $\ve{k}$ (\textit{i.e.}, two) while $\uv_{\nu\firstn{\ve{k}}}$ has the same dimension as $\ve{x}$. Sums are taken over all the accessible values of $\nu$ and $\ve{k}$.

The corresponding rate functions are
\begin{align}
  \qp{\ve{K}}{\ve{x}} & \! = \! \beta x_{S\equivn{\ve{K}}} \! \Bigl(\! \smallsum{\ve{k}} K_{I\firstn{\ve{k}}} \! \Bigr) \prob{S\secn{\ve{K}}}{S\equivn{\ve{K}},\ve{x},Y} \label{eq:SISfirst:qp} \\
  \qm{\ve{K}}{\ve{x}} & \! = \! \alpha\, x_{I\equivn{\ve{K}}}\, \prob{I\secn{\ve{K}}}{I\equivn{\ve{K}},\ve{x},Y} \quad . \label{eq:SISfirst:qm}
\end{align}
Note that, unlike \eqref{eq:SISpairRegular:qp}--\eqref{eq:SISpairRegular:qm} and \eqref{eq:SISpairER:qp}--\eqref{eq:SISpairER:qm}, the inference terms in \eqref{eq:SISfirst:qp}--\eqref{eq:SISfirst:qm} have the same form: the probability for a motif to be a $\nu\secn{\ve{K}}$ knowing (in addition to $\ve{x}$ and $Y$) that its central node is also the central node of a $\nu\equivn{\ve{K}}$ motif. Again assuming a large network size, they are provided by a product of multinomial distributions
\begin{equation}
  \prob{\nu\secn{\ve{K}}}{\nu\equivn{\ve{K}},\ve{x},Y} = \prod_{\nu'} \Bigl( \smallsum{\ve{k}} K_{\nu'\firstn{\ve{k}}} \Bigr)! \prod_{\ve{k}} \frac{1}{(K_{\nu'\firstn{\ve{k}}})!} \Biggl( \frac{ (k_\nu + 1) x_{\nu'\firstn{\ve{k} + \uv_\nu}} }{ \smallsum{\ve{k}'}\, k'_\nu x_{\nu'}\firstn{\ve{k}'} } \Biggr)^{\!\!K_{\nu'\firstn{\ve{k}}}} \label{eq:SISfirst:inference}
\end{equation}
\end{subequations}
which complete the rules $W$ for the first neighbourhood SIS model.

\begin{figure}
\begin{center}
\includegraphics{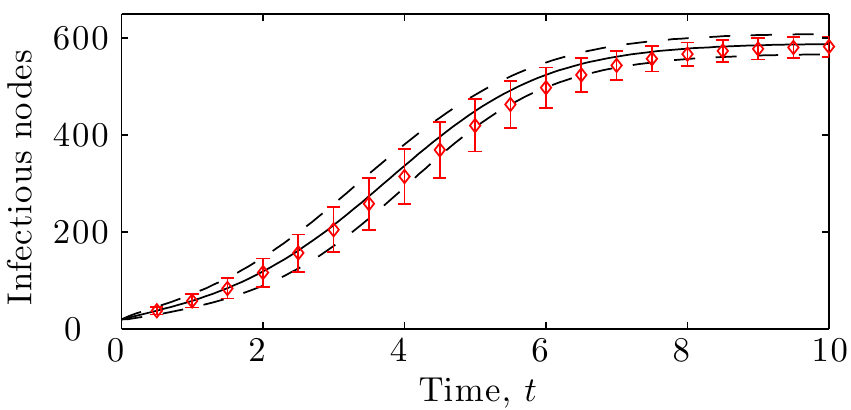}
\end{center}
\caption{(Online version in colour.) Time evolution of the number of infectious nodes for SIS dynamics ($\alpha = 1$ and $\beta = 1$) on a CM network for which the number of nodes of each degree is prescribed by the sequence $\{ 0, 50, 200, 450, 300 \}$ (total $N = 10^3$ nodes) with $2\%$ of the nodes of each degree initially infectious. The mean and range of one standard deviation above and bellow the mean are shown. Curves: simplified system. Symbols: full system ($10^5$ simulations). \label{fig:SISfirst:time}}
\end{figure}

Figure~\ref{fig:SISfirst:time} compares the results produced by this simplified model and the full one. Note that this is a stochastic version of the model presented in \cite{Marceau2010pre}, except that the network structure is here static.
%
\subsection{First neighbourhood SIR model \label{subsection:SIRfirst}}
As in Sec.~\ref{subsection:SISfirst}, we consider a full network model where the network structure is specified solely by the degree of its nodes. However, this time we consider SIR epidemiological dynamics: the accessible node states are $\nu \in \{ S, I, R \}$, infection is the same as in SIS but recovery is replaced by removal (see the introduction of Sec.~\ref{section:application} for details).

We define the forward transition event of type $\mathcal{I}\ve{K}$ to be the $\mathcal{I}$nfection of the central node of a $S\secn{\ve{K}}$ motif while a forward transition event of type $\mathcal{R}\ve{K}$ is the $\mathcal{R}$emoval of the central node of a $I\secn{\ve{K}}$ motif. There is no backward transition events. The model is specified by
\begin{subequations} \label{eq:SIRfirst}
\begin{align}
  \ve{x} & = \left( \cdots, x_{S\firstn{\ve{k}}}, \cdots, x_{I\firstn{\ve{k}}}, \cdots, x_{R\firstn{\ve{k}}}, \cdots \right) \\
  \ve{r}^{\mathcal{I}\ve{K}} & = \uv_{I\equivn{\ve{K}}} - \uv_{S\equivn{\ve{K}}} + \sum_\nu \sum_{\ve{k}} K_{\nu\firstn{\ve{k}}} \bigl( \uv_{\nu\firstn{\ve{k} + \uv_I}} - \uv_{\nu\firstn{\ve{k} + \uv_S}} \bigr) \\
  \ve{r}^{\mathcal{R}\ve{K}} & = \uv_{R\equivn{\ve{K}}} - \uv_{I\equivn{\ve{K}}} + \sum_\nu \sum_{\ve{k}} K_{\nu\firstn{\ve{k}}} \bigl( \uv_{\nu\firstn{\ve{k} + \uv_R}} - \uv_{\nu\firstn{\ve{k} + \uv_I}} \bigr)
\end{align}
\vspace{-\bigskipamount}
\begin{align}
  \qp{\mathcal{I}\ve{K}}{\ve{x}} & \! = \! \beta x_{\!S\equivn{\ve{K}}} \! \Bigl(\! \smallsum{\ve{k}} K_{I\firstn{\ve{k}}} \! \Bigr) \prob{S\secn{\ve{K}}}{S\equivn{\ve{K}},\ve{x},Y} \\
  \qp{\mathcal{R}\ve{K}}{\ve{x}} & \! = \! \alpha\, x_{I\equivn{\ve{K}}}\, \prob{I\secn{\ve{K}}}{I\equivn{\ve{K}},\ve{x},Y} \\
  \qm{\mathcal{I}\ve{K}}{\ve{x}} & \! = \! \qm{R\ve{K}}{\ve{x}} \! = \! 0 \label{eq:SIRfirst:qm}
\end{align}
\end{subequations}
where the inference terms are the same as in \eqref{eq:SISfirst:inference}.
%
\subsection{First neighbourhood on-the-fly SIR model \label{subsection:SIRotf}}
We take a different perspective to the problem considered in Sec.~\ref{subsection:SIRfirst} which requires to track much less elements in the state vector. Instead of considering ``complete'' first neighbourhood motifs, such as $\nu\firstn{\ve{k}}$, that specify the state of each of the central node's neighbours, we define the $\nu\otfn{\kappa}$ motif as a central node of state $\nu$ for which we know that it has $\kappa$ neighbours \emph{unknown to us}. This last statement is important: were we to learn the state of one of these neighbours, this would cease to be a $\nu\otfn{\kappa}$ motif and instead become a $\nu\otfn{\kappa-1}$ one. As usual, the state vector tracks the number of such motifs
\begin{subequations} \label{eq:SIRotf}
\begin{equation}
  \ve{x} = \bigl( \cdots, x_{S\otfn{\kappa}}, \cdots, x_{I\otfn{\kappa}}, \cdots, x_{R\otfn{\kappa}}, \cdots \bigr) \ .  \label{eq:SIRotf:x}
\end{equation}

We recall from Sec.~\ref{subsection:SISfirst} how a CM network is built in a computer simulation: for each $\kappa$, $n_\kappa$ nodes with $\kappa$ stubs are created and the stubs are then randomly paired to form links. From this perspective, $\nu\otfn{\kappa}$ may be reinterpreted as a $\nu$ state node with $\kappa$ unpaired stubs: as stubs are removed once they are paired in the computer simulation, neighbours that were unknown are removed from these motifs once they become known to us. Hence,
\begin{equation*}
\frac{\kappa x_{\nu\otfn{\kappa}} - \delta_{\nu\nu'} \delta_{\kappa\kappa'}}{\smallsum{\nu''} \smallsum{\kappa''} \, \kappa'' x_{\nu''\!\otfn{\kappa''}} - 1}
\end{equation*}
\emph{exactly} gives the probability for an unknown neighbours of the central node of $\nu'\otfn{\kappa'}$ to be the central node of $\nu\otfn{\kappa}$. Note that the Kronecker deltas \tbiglp $\delta_{ii'} = \smash{\bigl\{ \begin{matrix} \scriptstyle 1 & \scriptstyle i = i' \\[-1ex] \scriptstyle 0 & \scriptstyle i \neq i' \end{matrix} \bigr.}$ \tbigrp\ in the numerator and the $-1$ in the denominator both account for the stub of $\nu'\otfn{\kappa'}$ that we are pairing with a random stub.

A typical computer simulation would first build the network and then perform the SIR propagation dynamics on this network. However, we do not want to have to store the network structure for later consultation, which would require additional space in $\ve{x}$. Instead, we delay the network construction, leaving the stubs unpaired, and start the propagation dynamics right away. Just when the state of an unknown neighbour is required do we pair the corresponding stub with a randomly selected one, hence building the network \emph{on-the-fly}. Since the knowledge of stubs being matched will be lost in the future, this information must only be required at the very moment it is obtained if we want the resulting dynamics to \emph{exactly} reproduce the behaviour of the full system.

We thus take a different, although equivalent, perspective on the infection dynamics where each link is ``probed'' at most once. Instead of considering a probability $\beta\, dt$ of infection for each \emph{susceptible} neighbours of infectious nodes, we consider the same probability for each of their \emph{unknown} neighbours. Only when this probability returns true do we wonder about the state of the neighbour, whose state changes to infectious if and only if it was previously susceptible. In any case, we learned who were the neighbours of two nodes (\textit{i.e.}, the infectious and its neighbour) and we must update the state vector accordingly.

Hence, we define the $\mathcal{I}$nfection transition event $\mathcal{I}\nu\kappa\kappa'$ such that an infectious at the center of a $I\otfn{\kappa'}$ motif attempts to infect the $\nu$-state node at the center of a $\nu\otfn{\kappa}$ motif. Of course, only $\mathcal{I}S\kappa\kappa'$ transition events result in real infections. The more traditional transition event $\mathcal{R}\kappa$ corresponds to the $\mathcal{R}$emoval of the infectious node at the center of a $I\otfn{\kappa}$ motif, thus becoming $R\otfn{\kappa}$. The model is specified by
\begin{align}
  \ve{r}^{\mathcal{I}S\kappa\kappa'} \!\! & = \uv_{I\otfn{\kappa-1}} \! - \uv_{S\otfn{\kappa}} \! + \uv_{I\otfn{\kappa'-1}} \! - \uv_{I\otfn{\kappa'}} \label{eq:SIRotf:rIS} \\
  \ve{r}^{\mathcal{I}I\kappa\kappa'} \!\! & = \uv_{I\otfn{\kappa-1}} \! - \uv_{I\otfn{\kappa}} \! + \uv_{I\otfn{\kappa'-1}} \! - \uv_{I\otfn{\kappa'}} \label{eq:SIRotf:rII} \\
  \! \ve{r}^{\mathcal{I}R\kappa\kappa'} \!\! & = \uv_{R\otfn{\kappa-1}} \! - \uv_{R\otfn{\kappa}} \! + \uv_{I\otfn{\kappa'-1}} \! - \uv_{I\otfn{\kappa'}} \label{eq:SIRotf:rIR} \\
  \ve{r}^{\mathcal{R}\kappa} \! & = \uv_{R\otfn{\kappa}} \! - \uv_{I\otfn{\kappa}} \label{eq:SIRotf:rR}
\end{align}
\vspace{-1.5\bigskipamount}
\begin{align}
  \qp{\mathcal{I}\nu\kappa\kappa'}{\ve{x}} & = \beta\, \kappa'\, x_{I\otfn{\kappa'}}
  \frac{\kappa x_{\nu\otfn{\kappa}} - \delta_{I\nu} \delta_{\kappa\kappa'}}{\smallsum{\nu''} \smallsum{\kappa''} \, \kappa'' x_{\nu''\!\otfn{\kappa''}} - 1} \label{eq:SIRotf:qpI} \\
  \qp{\mathcal{R}\kappa}{\ve{x}} & = \alpha x_{I\otfn{\kappa}} \label{eq:SIRotf:qpR} \\
  \qm{\mathcal{I}\nu\kappa\kappa'}{\ve{x}} & = \qm{\mathcal{R}\kappa}{\ve{x}} = 0 \quad . \label{eq:SIRotf:qm}
\end{align}
\end{subequations}

\begin{figure}
\begin{center}
\includegraphics{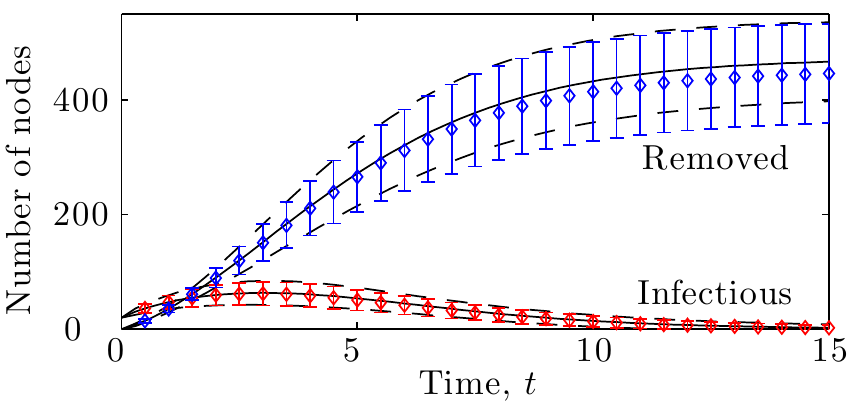}
\end{center}
\caption{(Online version in colour.) Time evolution of the number of infectious and removed nodes for SIR dynamics ($\alpha = 1$ and $\beta = 1$) on a CM network for which the number of nodes of each degree is prescribed by the sequence $\{ 0, 50, 200, 450, 300 \}$ (total $N = 10^3$ nodes) with $2\%$ of the nodes of each degree initially infectious (all others are susceptible). The mean and range of one standard deviation above and bellow the mean are shown. Curves: simplified system. Symbols: full system ($10^5$ simulations). \label{fig:SIRotf:time}}
\end{figure}
\begin{figure}
\begin{center}
\includegraphics{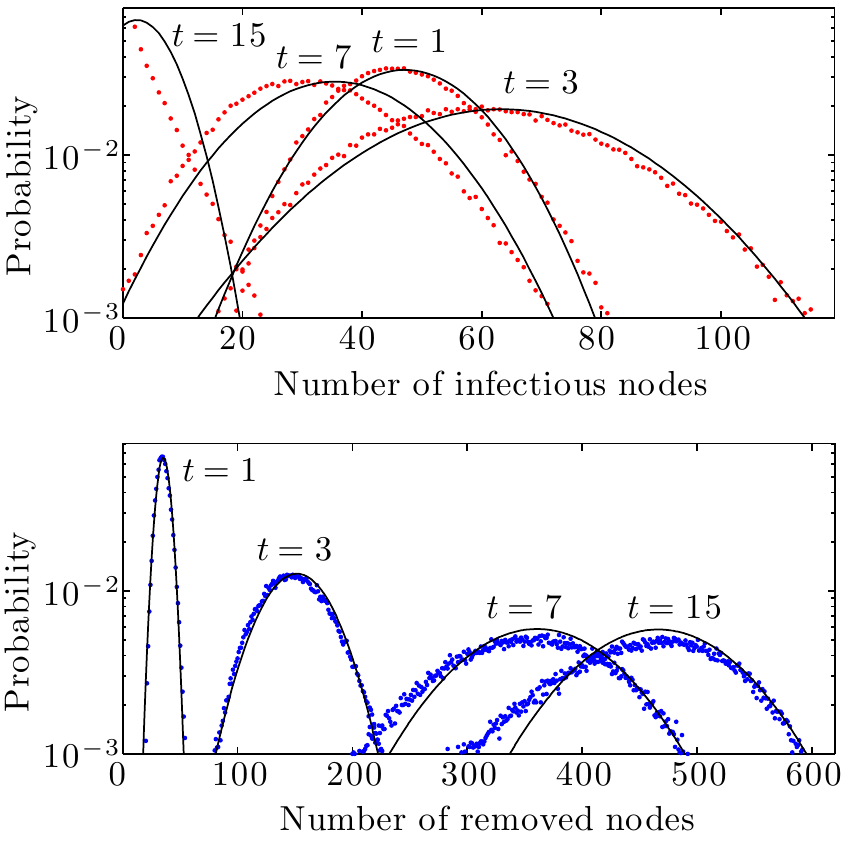}
\end{center}
\caption{(Online version in colour.) Probability distribution at different times for the number of infectious and removed nodes. The parameters are the same as in Fig.~\ref{fig:SIRotf:time}. Curves: simplified system. Symbols: full system ($10^5$ simulations). \label{fig:SIRotf:timeslices}}
\end{figure}

The system \eqref{eq:SIRotf} \emph{exactly} reproduces the behaviour of the full system through the solution of \eqref{eq:master}. Since \eqref{eq:ApproxMean}--\eqref{eq:GaussianApprox} are only approximations of \eqref{eq:master}, results obtained through these relationships are only approximative (Fig.~\ref{fig:SIRotf:time} and Fig.~\ref{fig:SIRotf:timeslices}). This model may be solved analytically for the mean value (see ESM \cite{noel11markovESM} \esmotfdeterministic) and the results are in agreement with \cite{volz08_jmathbio,miller09volz}. Moreover, ESM \cite{noel11markovESM} \esmotfalternative\ shows how \eqref{eq:SIRotf} may be rewritten with a state vector two thirds the size of \eqref{eq:SIRotf:x}. This is a generalization to the case $\alpha \neq 0$ of the model presented in \cite{noel11otf}. Further details are discussed in Section~\ref{subsection:discussion:otf}. We note that a conceptually similar approach has recently been developed independently \cite{decreusefond2011} as a tool for a mathematically rigorous proof that a specific heterogeneous mean field model \cite{volz08_jmathbio} holds in the limit of large network size.
%
\section{Discussion \label{section:discussion}}
We now take a retrospective look at the results presented in Sec.~\ref{section:application} and obtain from these special examples general considerations concerning our modelling approach.
%
\subsection{Accuracy of the results \label{subsection:discussion:accuracy}}
One of the aims of this paper is to obtain simplified models that accurately reproduce the behaviour of complex systems. Since approximations are usually involved, it is to be expected that the results of the simplified model only agree with those of the full system over some range of parameters, where the approximations were valid.

The parameters used in Fig.~\ref{fig:SISpairRegular:time}--\ref{fig:SIRotf:timeslices} were chosen in order to investigate the limits of our approximations: while there is no perfect correspondence between the results of the full and simplified systems, their agreement is probably sufficient for both qualitative and quantitative applications. We distinguish between two categories of approximations: those inherent to the use of \eqref{eq:ApproxMean}--\eqref{eq:GaussianApprox} and those due to the imperfect representation of $Z$ through $\ve{x}$ and $Y$.
%
\subsubsection{Gaussian approximation}
Since \eqref{eq:master} and \eqref{eq:SIRotf} define a system that \emph{exactly} reproduces the behaviour of the corresponding full system, any discrepancy in Fig.~\ref{fig:SIRotf:time} must originate from the use of the Gaussian approximation \eqref{eq:ApproxMean}--\eqref{eq:GaussianApprox}. An important requirement for this approximation to be valid is that the size $N$ of the system must be large.

Figures~\ref{fig:SISpairRegular:time}--\ref{fig:SIRotf:timeslices} all use networks of size $N = 1000$. As a rule of thumb, we found that \eqref{eq:ApproxMean}--\eqref{eq:GaussianApprox} perform better for networks of at least a few hundred nodes, which is the case of many relevant real-world systems. Note that, for very small systems (tens of nodes), one could also directly and completely solve \eqref{eq:master}.

While a large network size $N$ is required to justify treating the elements of $\ve{x}$ as real numbers, other phenomena may affect the validity of this approximation. For example, when the initial conditions are such that there is a single infectious node, the continuous approximation fails at considering the probability for that node to recover (or to be removed) before transmitting the infection to one of its neighbours. Figures~\ref{fig:SISpairRegular:time}--\ref{fig:SIRotf:timeslices} circumvent this problem by using an initial condition with $20$ infectious nodes: the probability for all of them to recover (or to be removed) before transmitting the infection is very low.

It is worth noting that the plateaux seen on Fig.~\ref{fig:SISpairRegular:time} and \ref{fig:SISpairER:time}--\ref{fig:SIRotf:time} reflect different dynamical behaviours for the SIS and SIR systems. Indeed, while the total number of removed nodes reaches a maximum in the SIR system because there are no infectious left to recover, the steady state observed at the later times for our SIS models corresponds to a constant flow of recovery and new infections. In the former case, the approximation errors performed at earlier times accumulate. In the later, the exact path taken to attain equilibrium is of lesser importance and errors do not accumulate the same way.
%
\subsubsection{Representation approximation \label{subsection:discussion:representation}}
In general, the simplified system will not exactly reproduce the behaviour of the full system, even when using \eqref{eq:master} instead of \eqref{eq:ApproxMean}--\eqref{eq:GaussianApprox}. This is the case of all our SIS models; while some of the discrepancy seen in Fig.~\ref{fig:SISpairRegular:time}--\ref{fig:SISfirst:time} is explained by the Gaussian approximation, the imperfect representation of $Z$ also contributes to the error.

Part of the problem can be understood as our failure to consider the correlation between the neighbours of a node and the time elapsed since this node has been in its present intrinsic state. For example, the neighbours of a susceptible node that has just recovered (\textit{i.e.}, it was infectious a moment ago) may be much different than those of a susceptible node that has recovered a long time ago, while being similar to those of a node that is still infectious. Hence, one could hope to improve these SIS models through changes in $Y$ alone (\textit{i.e.}, with the same $\ve{x}$): first estimate the probability distribution for the time since when each node has last changed state and then infer the neighbourhoods accordingly. An alternative that could be simpler to implement, at the cost of increasing the size of $\ve{x}$, would consist in tracking more exhaustive motifs (\textit{e.g.}, second neighbourhoods instead of first ones in Sec.~\ref{subsection:SISfirst}).

However, there are more intricate consequences to the recovery of infectious nodes on a structure that is fixed in time: if at some point all the nodes of the same component (\textit{i.e.}, a connected subnetwork that is disconnected from the rest of the network) are susceptible at the same time, then none of them may ever become infectious again. The connectivity of a network is strongly affected by the average degree of its nodes: our parameters correspond to an average degree of $5$ for Fig.~\ref{fig:SISpairRegular:time}--\ref{fig:SISpairER:time} (average degree of a neighbour also $5$) and of $3$ for Fig.~\ref{fig:SISfirst:time} (average degree of a neighbour $\approx 3.23$). When using smaller parameter values, this components-induced discrepancy becomes much larger since the simplified model then overestimates the number of infectious nodes. One could take the components into account by solving independent systems for each component (and merge the results afterwards) or by a clever adaptation of the inference process (see Sec.~\ref{subsection:discussion:otherinference} for possible directions). Note that these effects are usually much less important when the network structure changes over time.
%
\subsection{Pair-based models \label{subsection:discussion:pair}}
Compared to the other models presented in Sec.~\ref{section:application}, the two pair-based models of Sec.~\ref{subsection:pairSIS} use very small state vectors (\textit{i.e.}, two or three elements). This is an important advantage of pair-based models in general: there are usually much less pair and node motifs than, \textit{e.g.}, first neighbourhood ones, and tracking them thus requires much smaller $\ve{x}$. 

Although we limited our study of pair-based models to regular and Erd\H{o}s-R\'enyi networks, more complex network structures could also be considered. In the same way that \eqref{eq:SISpairRegular} and \eqref{eq:SISpairER} differ mostly by their inference terms, obtaining good inference from the little information stored in $\ve{x}$ is probably the principal challenge behind general and accurate pair-based stochastic models.

However, non-stochastic pair-based models are already possible on nontrivial network structures for SIR dynamics or, more generally, for processes such that a change in the state of one neighbour of a node can be treated as independent of that of another neighbour (SIS fails this assumption) \cite{millervolz11part1}. Knowing (in $Y$) that a system behaves in this manner greatly simplifies the inference process, and this is the main reason for the success of the SIR pair-based model for the evolution of mean values on CM networks that is presented in \cite{volz08_jmathbio,miller09volz}. Whether or not the same approach may be used to obtain stochastic results is an open question.
%
\subsection{First (and higher) neighbourhood models \label{subsection:discussion:first}}
By opposition, sufficiently accurate inference terms for first neighbourhood models are often straightforward to obtain. Although \eqref{eq:SISfirst:inference} may be difficult to appreciate at first sight, it is the only inference term used in both Sec.~\ref{subsection:SISfirst} and Sec.~\ref{subsection:SIRfirst}. In fact, \eqref{eq:SISfirst:inference} may well be the only inference term needed for generic first neighbourhood models for CM network structures.

Although first neighbourhood motifs are a ``natural language'' for expressing dynamics taking place on CM networks, they could also be used in the presence of other complex structures. This may be done through changes in $\ve{x}$ and/or $Y$; see ESM \cite{noel11markovESM} \esmmotif\ for details.

The generality and ease of design of first-neighbourhood models comes at a cost: the state vector $\ve{x}$ is typically much larger than it would be in an equivalent pair-model. How large is $\ve{x}$ strongly depends on the maximal node degree present in the network and on the total number of accessible intrinsic node states (see ESM \cite{noel11markovESM} \esmmotif\ for details). For typical values of these quantities, this does not cause major problems for the evaluation of the mean: numerically solving \eqref{eq:ApproxMean:dmudt} requires an acceptable amount of resources even for an $\ve{x}$ of dimension $10^6$ and \eqref{eq:ApproxMean:a} may often be simplified (\textit{i.e.}, summed analytically).

However, evaluating the covariance matrix using \eqref{eq:GaussianApprox} may cause problems: unless analytical simplifications are possible, solving this system scales as the square of the number of elements in $\ve{x}$. Future developments may decrease this bottleneck effect of the covariance matrix; see Sec.~\ref{subsection:discussion:statevsassumptions} for details. In any case, the size of $\ve{x}$ may be decreased by ``coarse graining'' the number of links between the central node and its neighbours; see ESM \cite{noel11markovESM} \esmmotifcoarsegraining\ for details.
%
\subsection{On-the-fly models \label{subsection:discussion:otf}}
The on-the-fly model presented in Sec.~\ref{subsection:SIRotf} for SIR dynamics on CM networks \emph{exactly} reproduces the behaviour of the full system. This is even more remarkable when one considers that the size of the state vector in the on-the-fly model is much smaller than in the alternative first neighbourhood model of Sec.~\ref{subsection:SIRfirst}. The reasons behind the success of the on-the-fly approach are similar to those discussed in Sec.~\ref{subsection:discussion:pair} for the pair models presented in \cite{millervolz11part1,volz08_jmathbio,miller09volz}: it is encoded in $Y$ that, for each link, we \emph{at most once} need to simultaneously know the state of the two nodes joined by that link \cite{noel11otf}.

The inference term \eqref{eq:SISfirst:inference} is of ``general purpose'' in the sense that its $Y$ does not provide information on the dynamical properties of the system, but only on how the motifs in $\ve{x}$ may be interconnected. This is why both \eqref{eq:SISfirst} and \eqref{eq:SIRfirst} rely on \eqref{eq:SISfirst:inference}.

However, the inference terms of \eqref{eq:SIRotf} have a specific character: $Y$ contains information about \eqref{eq:SIRotf} itself. Any change to the dynamics implies changes in the inference terms, with no guarantee that an acceptable solution exists. In fact, \eqref{eq:SIRotf} was \emph{designed} with this problem in mind. In other words, we obtained a simple and reliable model at the cost of ``pre-computations'' in the design process. Of all the possibilities in model-space, the information acquired by pointing at this specific one is what replaces the reduced size of the state vector. The same could be said of the deterministic SIR pair-based model on CM networks presented in \cite{volz08_jmathbio,miller09volz}.

By contrast with the case discussed in Sec.~\ref{subsection:discussion:first}, the small size of the state vector here allows for evaluations of the covariance matrix through \eqref{eq:GaussianApprox}, even when relatively high degree nodes are present. Alternatively, one may take advantage of the fact that, even for more complicated dynamics, the state vector of on-the-fly models can remain of manageable size for mean values calculations; see the introduction to ESM \cite{noel11markovESM} \esmmotif\ for the concrete example of \cite{Marceau2011pre}.
%
\subsection{Complicated states \textit{vs} complex assumptions \label{subsection:discussion:statevsassumptions}}
Section~\ref{subsection:discussion:otf} revealed an unexpected depth to $Y$: one may achieve models of similar levels of accuracy by trading off complexity in the assumptions for a reduction in the size of the state vector $\ve{x}$. As an extreme example, if $Y$ already gives the full behaviour of the system, then there is no need for tracking any information in $\ve{x}$. Without reaching such extremes, our on-the-fly model and the deterministic SIR pair-based model presented in \cite{volz08_jmathbio,miller09volz} both demonstrate the benefits of investing some time in the assumptions of our models.

While these examples required case-by-case analysis, one may benefit from the same realization in a general context: a first simplified model ($W$, $\ve{x}$ and $Y$) may generate the assumptions $Y'$ to a different simplified model ($W'$, $\ve{x}'$ and $Y'$). For example, when some dynamical process (\textit{e.g.}, SIS or SIR) occurs on a network whose structure changes in time independently from this dynamics, one could obtain a first model for the structure alone and then feed the results to the second model, handling the remaining dynamics. Even more generally, one could compensate for the higher computational requirements of \eqref{eq:GaussianApprox} by first solving \eqref{eq:ApproxMean} on an elaborate model then feeding the resulting mean values to a simpler model for the sole purpose of estimating the covariance matrix.
%
\subsection{Additional inference tools \label{subsection:discussion:otherinference}}
While we introduced $Y$ as a direct application of Bayes' rule, we have now seen that useful assumptions may be obtained by other means, including the solution of another system of the form \eqref{eq:master}. The next step in this direction would be to improve our inference process using alternative tools and models available to network science.

For example, branching processes \cite{newman01_pre} may be used to infer information concerning the connectivity and the components of the network structure. As discussed in Sec.~\ref{subsection:discussion:representation}, this point was a major shortfall of SIS models. This approach is even more interesting for the recently developed tools \cite{allard09_pre,karrer10} that are particularly compatible with the motifs and intrinsic node state approach presented in this paper.

Another tool of considerable interest are exponential random networks \cite{park04a_pre}. Indeed, these maximum entropy methods can simplify inferences that would have otherwise been prohibitively complex. Once again, this approach may be generalized to different kind of motifs and intrinsic node states.
%
\section{Conclusion: general applicability \label{conclusion}}
Although the examples of Sec.~\ref{section:application} focus on SIS and SIR dynamics, any specificity that could be modelled through a standard epidemiological compartmental model may \textit{a priori} be considered by our approach: genders, age groups, vaccination, incubation period, disease phases, \textit{etc}. Each compartment simply becomes an accessible node state in our formalism; see ESM \cite{noel11markovESM} \esmfullstate\ for details.

Furthermore, population dynamics considerations may be accounted for in a straightforward manner. Assuming first neighbourhood motifs, births and deaths of individuals correspond to events adding and removing motifs, respectively. Similarly, changes in interaction patterns amount to events replacing the affected motifs by new ones. In fact, from the model's perspective, there is no important distinction between a change in the interaction structure of the population and a change in the node states: both are events affecting motifs.

The generality of our systematic approach and the fact that its assumptions are explicitly stated suggests that it could be used as a common ground for comparing existing models too complex for direct comparison. Indeed, by considering such an existing model as the full system (specified by $V_1$ and $Z_1$), one may seek a simplified system (specified by $W_1$, $X_1$ and $Y_1$) approximately reproducing the original model (over a sufficient range of parameters).

If some transition event (in $W_1$) appears essential, this may reveal an important feature of the original model; the same holds true for motifs (in $X_1$) and prior knowledge (in $Y_1$). Moreover, assuming that this procedure has been done for a second existing model (specified by $V_2$ and $Z_2$), one may directly compare their simplified version in a common framework, which will help identify the assumptions required for their description. Note that this perspective is similar to a commutation diagram
\begin{equation*}
\begin{matrix}
(V_1,Z_1) & \overset{\mathclap{\text{Difficult to compare}}}{\mathclap{\longleftrightarrow}\mathclap{/}} & (V_2,Z_2) \\[1ex]
\mathllap{\scriptstyle\text{Equivalent behaviour\ }}\Updownarrow & & \Updownarrow \mathrlap{\scriptstyle\text{\ Equivalent behaviour}} \\
(W_1,X_1,Y_1) & \overset{\text{Comparable}}{\mathclap{\longleftrightarrow}\mathclap{\phantom{/}}} & (W_2,X_2,Y_2) \mathrlap{\qquad .}
\end{matrix}
\end{equation*}
For example, if $X_1 = X_2$ and $Y_1 = Y_2$, we know that the discrepancies between the two original models is imputable to the difference in the transition events. Finding a minimal set of changes to $W_1$ and/or $W_2$ causing both models to agree may then help identify the very cause of the discrepancies.
%
\section*{Acknowledgements}
The research team is grateful to the Canadian Institutes of Health Research (CIHR), the Natural Sciences and Engineering Research Council of Canada (NSERC) and the Fonds de recherche du Qu\'ebec --- Nature et technologies (FRQ---NT) for financial support.
%
%
\bibliographystyle{vancouver}
%

%


\section*{Supplementary material}

# Electronic Supplementary Material to "Epidemics on contact networks: a general stochastic approach"

Pierre-André Noël, Antoine Allard, Laurent Hébert-Dufresne, Vincent Marceau and Louis J. Dubé
Département de Physique, de Génie Physique et d'Optique, Université Laval, Québec (QC), Canada

December 8, 2011

**Abstract**

This document provides supplementary informations to "Markov processes on complex networks with applications to spreading dynamics" [1], which will hereafter be referred to as the "main text". Equation numbers from the main text are preceded by the letters "MT".

## 1 State of the full system ($Z$)

The state of the full system is specified by the intrinsic state of all of its components and the network structure governing their interactions. As defined in the main text [1] §II A, a node has at any time a single intrinsic node state $\nu$. We note $\mathcal{N}$ the total number of accessible node states, which takes its minimal value $\mathcal{N} = 1$ when nodes are intrinsically indiscernible. In addition, we introduce the *intrinsic link state* $\ell$. In a similar way, $\mathcal{L}$ is the total number of accessible link states, which takes its minimal value $\mathcal{L} = 1$ for intrinsically indiscernible links ($\mathcal{L} = 1$ throughout the main text). Examples of link states may include any relevant characteristics of the interactions: their context (*e.g.*, professional, friendship, partnership, ...), their dynamical status (*e.g.*, active or inactive), their weight (*e.g.*, strong, normal, weak, ...), *etc.*

While each node (or link) may be in only one intrinsic state at any given time, different pieces of information may be encoded in this single state. As it is the case in standard compartmental models, a simple Cartesian product may suffice to this task. For example, an SIR dynamics (three epidemiological states, see main text [1] §III) where we discern two genders (male and female) would result in a total of $\mathcal{N} = 6$ accessible node states. Note that there is no particular problem caused by the fact that one of these characteristics is susceptible to change during the process while the other one remains constant.

In addition to intrinsic link states, links may (or may not) be directed. The links of the networks discussed in the main text [1] are *undirected*: the interaction between the two linked nodes is bidirectional and symmetric. *Directed* links represent interactions that are either unidirectional or asymmetrical. A network that has only undirected links is said to be *undirected*, one that has only directed links is *directed* and one that has both is *semi-directed*.

## 2 More on motifs

We introduce new motifs and generalize those presented in the main text [1] for intrinsic link types and directed (or semi-directed) networks. Table 1 summarizes the total number of motifs in each of these classes.

Of high practical importance is the fact that the entries of this table differ widely in their scaling behaviours. For example, the on-the-fly model [2] for two interacting SIR dynamics each propagating on their own network structure uses $\mathcal{N} = 9$ (two SIR), $\mathcal{L} = 3$ (links of the first network alone, links of the second network alone and overlapping links) and $\mathcal{D} = 1$ (undirected network). Looking up in table 1, we see that this requires of the order of $\mathcal{K}^3$ on-the-fly (degreed node) motifs, where $\mathcal{K}$ denotes the highest accessible node degree. By opposition, implementing a first neighbourhood version of this model would require of the order of $\mathcal{K}^{27}$ motifs!

1

Table 1: Number of motifs in selected classes. We defined $\mathcal{D} = 1$ for undirected, $\mathcal{D} = 2$ for directed and $\mathcal{D} = 3$ for semi-directed networks. Moreover, $\mathcal{K}$ denotes the highest accessible node degree.

| Motifs | Number |
| --- | --- |
| Node | $\mathcal{N}$ |
| Pair | $\frac{1}{2}\mathcal{L}\mathcal{N}(\mathcal{D}\mathcal{N} + (\mathcal{D}-2)^2)$ |
| Triple | $\frac{1}{2}\mathcal{D}\mathcal{L}\mathcal{N}^2(\mathcal{D}\mathcal{L}\mathcal{N}+1)$ |
| Triangle | $\frac{1}{3}\mathcal{D}\mathcal{L}\mathcal{N} + \frac{1}{2}\mathcal{D}\big((\mathcal{D}-2)\mathcal{L}\mathcal{N}\big)^2 + \frac{1}{6}(\mathcal{D}\mathcal{L}\mathcal{N})^3$ |
| Degreed node | $\mathcal{N}\binom{\mathcal{K}+\mathcal{D}\mathcal{L}}{\mathcal{K}}$ |
| Degreed pair | $\frac{\mathcal{L}\mathcal{N}}{2}\binom{\mathcal{K}-1+\mathcal{D}\mathcal{L}}{\mathcal{K}-1}\Big(\mathcal{D}\mathcal{N}\binom{\mathcal{K}-1+\mathcal{D}\mathcal{L}}{\mathcal{K}-1}+(\mathcal{D}-2)^2\Big)$ |
| First neighbourhood (node) | $\mathcal{N}\binom{\mathcal{K}+\mathcal{D}\mathcal{L}\mathcal{N}}{\mathcal{K}}$ |
| First neighbourhood pair | $\frac{\mathcal{L}\mathcal{N}}{2}\binom{\mathcal{K}-1+\mathcal{D}\mathcal{L}\mathcal{N}}{\mathcal{K}-1}\Big(\mathcal{D}\mathcal{N}\binom{\mathcal{K}-1+\mathcal{D}\mathcal{L}\mathcal{N}}{\mathcal{K}-1}+(\mathcal{D}-2)^2\Big)$ |
| Second neighbourhood (node) | $\mathcal{N}\binom{\mathcal{K}+\mathcal{D}\mathcal{L}\mathcal{N}\binom{\mathcal{K}-1+\mathcal{D}\mathcal{L}\mathcal{N}}{\mathcal{K}-1}}{\mathcal{K}}$ |

## 2.1 Pair motifs

We note $\nu \overset{\ell}{-} \nu'$ (resp. $\nu \overset{\ell}{\to} \nu'$) an undirected (resp. directed) pair motif formed of a state $\nu$ node linked through a state $\ell$ link to a state $\nu'$ node. In the case of directed motifs, the direction of the arrow usually specifies the "strongest causal effect" of this asymmetric interaction (although this needs not be the case). All the undirected and directed motifs are possible in a semi-directed network. We may omit the index $\ell$ over the links when $\mathcal{L} = 1$.

## 2.2 Triple motifs

We note $\nu \overset{\ell}{-} \nu' \overset{\ell'}{-} \nu''$ an undirected triple motif formed of a state $\nu$ node linked through a state $\ell$ link to a state $\nu'$ node itself linked through a state $\ell'$ link to a state $\nu''$ node. As for pair motifs, each of these nodes may have other neighbours than those that are explicitly specified. The notation directly generalizes to directed (*e.g.*, $\nu \overset{\ell}{\to} \nu' \overset{\ell'}{\leftarrow} \nu''$) and semi-directed (*e.g.*, $\nu \overset{\ell}{-} \nu' \overset{\ell'}{\to} \nu''$) triple motifs.

The term "2-star" is often used to refer to a triple motif for which both extremities (*e.g.*, the nodes of state $\nu$ and $\nu''$ in the motif $\nu \overset{\ell}{-} \nu' \overset{\ell'}{-} \nu''$) are explicitly *forbidden* to be neighbours. In models that also use triangle motifs, 2-star motifs may explicit the absence of the last link that would form a triangle. Another common use of triple motifs comes in the inference process of (usually deterministic) pair based models.

## 2.3 Triangle motifs and other small subnetworks

Three nodes that are all neighbours of each other form a triangle motif. An horizontal bracket represents the additional link that would be missing in the analogous triple motif, *e.g.*,

$$\underset{\ell''}{\underbrace{\nu \overset{\ell}{-} \nu' \overset{\ell'}{-} \nu''}}$$

for an undirected network.

Triangle motifs are usually considered in models that should account for clustering. Their number may either directly be tracked in the state vector **x** [3] or their implicit presence (stated in $Y$, *e.g.*, through a clustering coefficient) may be accounted for in the inference process [4].

The same notation may be generalized to other motifs consisting of small subnetworks, *e.g.*, square motifs [3].

## 2.4 Clique motifs

A vague definition of a clique motif is "a subgroup of nodes that share more links among themselves than what could be expected otherwise for the same number of randomly selected nodes". In applications, one may refine this definition according to the specificities of the problem at hand, *e.g.*, "an Erdős-Rényi subnetwork (link probability $p$) of $n_S$ susceptible nodes and $n_I$ infectious nodes". Clique motifs are usually considered in models that should account for community structure [5].

## 2.5 Degreed motifs

A degreed motif is a motif for which we know the degree of all the nodes forming the motif: a degreed node motif is a node of specified state and degree; a degreed link motif is two nodes of specified state and degree that are known to be neighbours; *etc*. The in-degree and out-degree are both specified in directed and semi-directed networks; the latter cases also specify undirected degrees. Likewise, degrees pertaining to different types of links are specified independently.

In the same way that pair motifs are usually combined with node motifs, degreed pair motifs are usually combined with degreed node motifs [6]. Note that the on-the-fly motifs $\nu\Lambda_1(\kappa)$ presented in the main text [1] §III D can be understood as a special case of degreed node motifs where the degree is replaced by the "degree to unknown nodes".

## 2.6 $n$-th neighbourhood motifs

Similarly to degreed motifs, an $n$-th neighbourhood motif is a motif for which we specify the state of all the $n$-th neighbours of the nodes forming the original motif. Hence, the notation $\underline{\nu}\Gamma_1(\mathbf{k})$ [resp. $\underline{\nu}\Gamma_2(\mathbf{K})$] of the main text corresponds to a first (resp. second) neighbourhood *node* motif. These concepts are directly generalizable to types of links and to directed or semi-directed networks.

First neighbourhood node motifs can be understood as tracking the correlation between the state of a node and the state of all its neighbours. By opposition, degreed pair motifs track the correlation between the state and degree of two neighbouring nodes. While similar information may be obtained from both motif classes, a model based on one may perform better than a model based on the other depending on the characteristics of the full system to be modelled.

## 2.7 Coarse-grained degree and/or neighbourhood

Not all entries of table 1 depend on the maximal degree $\mathcal{K}$, but those that do quickly increase for large $\mathcal{K}$. This is problematic since many real-world systems contain high degree nodes.

However, one may overcome this limitation by coarse-graining degrees through ranges: the range containing the degree of a node is specified instead of the degree itself. For example, given the ranges

$$\underbrace{[0,0]}_{\text{range 0}}, \underbrace{[1,1]}_{\text{range 1}}, \underbrace{[2,3]}_{\text{range 2}}, \underbrace{[4,7]}_{\text{range 3}}, \underbrace{[8,15]}_{\text{range 4}}, \underbrace{[16,31]}_{\text{range 5}} \text{ and } \underbrace{[32,63]}_{\text{range 6}},$$

we would say of a degree 23 node that its degree lies within range 5. Hence, for the purpose of evaluating the number of motifs in table 1, one should here use $\mathcal{K} = 6$ (one less than the total number of ranges) instead of $\mathcal{K} = 63$ (highest representable degree).

While the previous example used powers of 2 for simplicity, a slower increase is probably desirable in most applications. However, since the number of neighbourhood and degreed motifs strongly depends on $\mathcal{K}$, even the slightest reduction in this number may be significant. Note that this coarse-graining method is of particular interest when the real-world data used to calibrate the model is already coarse-grained, which is commonly the case for census data.

# 3 Deterministic solution of on-the-fly SIR model

This section provides the deterministic solution of (MT10) from the main text [1]. Our result is the same one as for the SIR pair model presented in [7, 8].

We first rewrite (MT3) for the specific case of (MT10)

$$\frac{d\boldsymbol{\mu}(t)}{dt} = \sum_{\nu}\sum_{\kappa}\sum_{\kappa'} \mathbf{r}^{\mathcal{I}\nu\kappa\kappa'} q^+_{\mathcal{I}\nu\kappa\kappa'}(\boldsymbol{\mu}(t), Y) + \sum_{\kappa} \mathbf{r}^{\mathcal{R}\kappa} q^+_{\mathcal{R}\kappa}(\boldsymbol{\mu}(t), Y) \tag{1}$$

then collect the contributions to $\mu_{S\Lambda_1(\kappa)}$, $\mu_{I\Lambda_1(\kappa)}$ and $\mu_{R\Lambda_1(\kappa)}$ (dropping the functional dependencies for brevity)

$$\frac{d\mu_{S\Lambda_1(\kappa)}}{dt} = -\sum_{\kappa'} q^+_{\mathcal{I}S\kappa\kappa'} \tag{2a}$$

$$\frac{d\mu_{I\Lambda_1(\kappa)}}{dt} = \sum_{\kappa'} \left( q^+_{\mathcal{I}S(\kappa+1)\kappa'} + q^+_{\mathcal{I}I(\kappa+1)\kappa'} - q^+_{\mathcal{I}I\kappa\kappa'} \right) + \sum_{\nu}\sum_{\kappa'} \left( q^+_{\mathcal{I}\nu\kappa'(\kappa+1)} - q^+_{\mathcal{I}I\kappa'\kappa} \right) - q^+_{\mathcal{R}\kappa} \tag{2b}$$

$$\frac{d\mu_{R\Lambda_1(\kappa)}}{dt} = \sum_{\kappa'} \left( q^+_{\mathcal{I}R(\kappa+1)\kappa'} - q^+_{\mathcal{I}R\kappa\kappa'} \right) + q^+_{\mathcal{R}\kappa} \quad . \tag{2c}$$

Using the definitions

$$\lambda = \sum_{\kappa} \kappa \mu_{I\Lambda_1(\kappa)} \quad \text{and} \quad \omega = \sum_{\nu}\sum_{\kappa} \kappa \mu_{\nu\Lambda_1(\kappa)} \tag{3}$$

where $\lambda$ is the total number of stubs belonging to infectious nodes and $\omega$ is the total number of stubs in the system, (2) becomes

$$\frac{d\mu_{S\Lambda_1(\kappa)}}{dt} = -\frac{\beta\lambda\kappa\mu_{S\Lambda_1(\kappa)}}{\omega} \tag{4a}$$

$$\frac{d\mu_{I\Lambda_1(\kappa)}}{dt} = \frac{\beta\lambda(\kappa+1)\mu_{S\Lambda_1(\kappa+1)}}{\omega} - \alpha\mu_{I\Lambda_1(\kappa)} + \beta\left(1 + \frac{\lambda}{\omega}\right)\left((\kappa+1)\mu_{I\Lambda_1(\kappa+1)} - \kappa\mu_{I\Lambda_1(\kappa)}\right) \tag{4b}$$

$$\frac{d\mu_{R\Lambda_1(\kappa)}}{dt} = \frac{\beta\lambda}{\omega}\left((\kappa+1)\mu_{R\Lambda_1(\kappa+1)} - \kappa\mu_{R\Lambda_1(\kappa)}\right) + \alpha\mu_{I\Lambda_1(\kappa)} \quad . \tag{4c}$$

Note that (MT10f) has been approximated by dropping the Kronecker delta in the numerator and the $-1$ in the denominator.

We now consider the evolution of the total number of stubs $\omega$ by summing the contributions from (4)

$$\frac{d\omega}{dt} = \sum_{\nu}\sum_{\kappa} \kappa \frac{d\mu_{\nu\Lambda_1(\kappa)}}{dt} = -2\beta\lambda \quad . \tag{5}$$

One may understand (5) as "during the time interval $[t, t+dt)$, each one of the $\lambda$ stubs belonging to infectious nodes have probability $\beta dt$ to be paired to another stub, thus causing a decrease by 2 of $\omega$. Noting $\omega_0 = \omega(0)$ the total number of stubs in the initial condition, we introduce the change of variable

$$\theta = \sqrt{\frac{\omega}{\omega_0}} \quad \text{such that} \quad \frac{d\theta}{dt} = -\frac{\beta\lambda}{\theta\omega_0} \tag{6}$$

Notice that $t = 0$ corresponds to $\theta = 1$ and that $\theta$ *decreases* with time. Using this change of variable in (4a) gives

$$\frac{d\mu_{S\Lambda_1(\kappa)}}{d\theta} = \frac{\kappa\mu_{S\Lambda_1(\kappa)}}{\theta} \tag{7}$$

which has the solution

$$\mu_{S\Lambda_1(\kappa)}(t) = x_{S\Lambda_1(\kappa)}(0)\left(\theta(t)\right)^{\kappa} \quad . \tag{8}$$

using the initial condition $\mu_{S\Lambda_1(\kappa)}(0) = x_{S\Lambda_1(\kappa)}(0)$.

For convenience, we define

$$f(\theta) = \sum_{\kappa} x_{S\Lambda_1(\kappa)}(0)\theta^{\kappa} \quad . \tag{9}$$

Noticing that

$$\sum_\kappa \kappa(\kappa+1)\mu_{S\Lambda_1(\kappa+1)} = \theta^2 \sum_\kappa (\kappa-1)\kappa x_{S\Lambda_1(\kappa)}(0)\theta^{\kappa-2} = \theta^2 f''(\theta) \quad, \tag{10}$$

we obtain the evolution of the total number $\lambda$ of stubs belonging to infectious nodes by summing the contributions (4b)

$$\frac{d\lambda}{dt} = \sum_\kappa \kappa \frac{d\mu_{I\Lambda_1(\kappa)}}{dt} = \frac{\beta\lambda\theta^2 f''(\theta)}{\omega} - \beta\lambda\left(1+\frac{\lambda}{\omega}\right) - \alpha\lambda \quad. \tag{11}$$

Again using the change of variable (6), we get

$$\frac{d\lambda}{d\theta} = \frac{\lambda}{\theta} + \theta\omega_0\left(1+\frac{\alpha}{\beta}\right) - \theta f''(\theta) \tag{12}$$

which has the solution

$$\lambda = \theta^2 \omega_0\left(1+\frac{\alpha}{\beta}\right) - \theta\omega_0\frac{\alpha}{\beta} - \theta f'(\theta) \tag{13}$$

for an initial condition without removed nodes $[i.e., \lambda(0) = \omega_0 - f'(1)]$.

Using this solution in (6) gives

$$\frac{d\theta}{dt} = -\beta\theta + \alpha(1-\theta) + \beta\frac{f'(\theta)}{\omega_0} \tag{14}$$

whose solution provides $\theta(t)$. Using

$$S(t) = f(\theta(t)) \tag{15a}$$

$$I(t) = N - S(t) - R(t) \tag{15b}$$

$$\frac{dR(t)}{dt} = \alpha I(t) \quad, \tag{15c}$$

we finally obtain the total number $S = \sum_\kappa \mu_{S\Lambda_1(\kappa)}$ of susceptible, $I = \sum_\kappa \mu_{I\Lambda_1(\kappa)}$ of infectious and $R = \sum_\kappa \mu_{R\Lambda_1(\kappa)}$ of removed nodes at any given time $t$. A direct application of (8)–(9) provides (15a), conservation of the nodes provides (15b) and using the definitions of $I(t)$ and $R(t)$ in (4c) provides (15c). Although obtained differently, this solution corresponds to that of the pair-based SIR model presented in [7, 8].

## 4 Alternative form of on-the-fly SIR model

The $R\Gamma_1(\mathbf{k})$ motifs were included in (MT10a) for the sake of clarity alone: noting $s$ an unmatched stub motif, we could rewrite (MT10) with a state vector two thirds the size of (MT10a)

$$\mathbf{x} = \left(\cdots, x_{S\Lambda_1(\kappa)}, \cdots, x_{I\Lambda_1(\kappa)}, \cdots, x_s\right)$$

$$\mathbf{r}^{\mathcal{I}S\kappa\kappa'} = \widehat{\mathbf{e}}_{I\Lambda_1(\kappa-1)} - \widehat{\mathbf{e}}_{S\Lambda_1(\kappa)} + \widehat{\mathbf{e}}_{I\Lambda_1(\kappa'-1)} - \widehat{\mathbf{e}}_{I\Lambda_1(\kappa')} - 2\widehat{\mathbf{e}}_s$$

$$\mathbf{r}^{\mathcal{I}I\kappa\kappa'} = \widehat{\mathbf{e}}_{I\Lambda_1(\kappa-1)} - \widehat{\mathbf{e}}_{I\Lambda_1(\kappa)} + \widehat{\mathbf{e}}_{I\Lambda_1(\kappa'-1)} - \widehat{\mathbf{e}}_{I\Lambda_1(\kappa')} - 2\widehat{\mathbf{e}}_s$$

$$\mathbf{r}^{\mathcal{I}R\kappa\kappa'} = \widehat{\mathbf{e}}_{I\Lambda_1(\kappa'-1)} - \widehat{\mathbf{e}}_{I\Lambda_1(\kappa')} - 2\widehat{\mathbf{e}}_s$$

$$\mathbf{r}^{\mathcal{R}\kappa} = -\widehat{\mathbf{e}}_{I\Lambda_1(\kappa)}$$

$$q^+_{\mathcal{I}S\kappa\kappa'}(\mathbf{x},Y) = \beta\kappa' x_{I\Lambda_1(\kappa')} \frac{\kappa x_{S\Lambda_1(\kappa)}}{x_s - 1}$$

$$q^+_{\mathcal{I}I\kappa\kappa'}(\mathbf{x},Y) = \beta\kappa' x_{I\Lambda_1(\kappa')} \frac{\kappa x_{I\Lambda_1(\kappa)} - \delta_{\kappa\kappa'}}{x_s - 1}$$

$$q^+_{\mathcal{I}R\kappa\kappa'}(\mathbf{x},Y) = \beta\kappa' x_{I\Lambda_1(\kappa')} \frac{x_s - 1 - \sum_{\kappa''}\kappa''\left(x_{S\Lambda_1(\kappa'')} + x_{I\Lambda_1(\kappa'')}\right)}{x_s - 1}$$

$$q^+_{\mathcal{R}\kappa}(\mathbf{x},Y) = \alpha x_{I\Lambda_1(\kappa)}$$

$$q^-_{\mathcal{I}\nu\kappa\kappa'}(\mathbf{x},Y) = q^-_{\mathcal{R}\kappa}(\mathbf{x},Y) = 0 \quad.$$

Note that $x_s$ plays the same role as $\omega$ in Sec. 3. Proceeding similarly, the state vector $\mathbf{x} = \left(\cdots, x_{S\Lambda_1(\kappa)}, \cdots, x_s\right)$ would suffice for a SI model (*i.e.*, $\alpha = 0$) [9].


\end{document}